\documentclass[prb,showpacs,twocolumn]{revtex4}
\usepackage{epsfig}
\usepackage{graphics}
\usepackage{graphicx}
\usepackage{dcolumn}
\usepackage{bm}

\newcommand{\sba}{\begin{subeqnarray}}
\newcommand{\sea}{\end{subeqnarray}}

\def\cm-1{cm$^{-1}$}

\begin{document}

\title{Single-site Anderson model. II Perturbation theory of symmetric model}
\author{V.\ A.\ Moskalenko$^{1,2}$}
\email{moskalen@thsun1.jinr.ru}
\author{P.\ Entel$^{3}$}
\author{D.\ F.\ Digor$^{1}$}\author{L.\ A.\ Dohotaru$^{4}$} \author{R.\ Citro$^{5}$}
\affiliation{$^{1}$Institute of Applied Physics, Moldova Academy
of Sciences, Chisinau 2028, Moldova} \affiliation{$^{2}$BLTP,
Joint Institute for Nuclear Research, 141980 Dubna, Russia}
\affiliation{$^{3}$University of Duisburg-Essen, 47048 Duisburg,
Germany}
\affiliation{$^{4}$Technical University, Chisinau 2004,
Moldova}
\affiliation{$^{4}$Dipartimento di Fisica "E. R. Caianiello", Universit\'{a} degli Studi
di Salerno and CNISM, Unit\'{a} di ricerca di Salerno, Via S. Allende, 84081 Baronissi (SA), Italy}\date{\today}

\begin{abstract}
%
The strong electron correlations caused by Coulomb interaction of
impurity electrons are taken into account. The infinite series of
diagrams containing irreducible Green's functions are summed. For
symmetric Anderson model we establish the antisymmetry property of
the impurity Green's function, formulate the exact Dyson type
equation for it, find the approximate correlation function
$Z_{\sigma }(i\omega $) and solve the integral equation which
determines the full propagator of the impurity electrons.
Analytical continuation of the obtained Matsubara Green's function
determines the retarded one and gives the possibility to find the
spectral function of impurity electrons. The existence of two
resonances of this function has been proved. The smooth behaviour
was found near the Fermi surface. The two resonances situated
symmetrical to the Fermi surface correspond to the energies of
quantum transitions of the impurity electrons. The widths and
heights of these resonances are established.
%
\end{abstract}
%
\maketitle



\section{Introduction}


We shall investigate the properties of the normal state of
single-site Anderson model. For that we shall use the results of
diagrammatic theory for this model developed in our previous
paper$^{[1]}$. In that investigation the notion of irreducible
Green's function has been introduced. These functions contain the
main spin, charge and pairing fluctuations caused by the strong
Coulomb repulsion of impurity electrons. We have determined the
notion of correlation function $Z_{\sigma \sigma ^{\prime }}$ as
composed from strong connected diagrams containing the irreducible
Green's functions. In superconducting state of the system there
are additional correlation functions $Y_{\sigma \sigma ^{\prime
}}$ and $\overline{Y}_{\overline{\sigma }\sigma ^{\prime }}$ which
are the order parameters of the system. The correlation functions
are the main elements of the Dyson type equations for
one--particle renormalized Green's function of conducting and
impurity electrons. We shall restrict ourselves only by the
discussion of the properties of the normal phase of the system and
determine the corresponding Green's functions of conduction and
impurity electrons:
%
\begin{eqnarray}
%
G_{\sigma \sigma ^{\prime }}(\mathbf{k,}\tau |\mathbf{k}^{\prime },\tau
^{\prime }) &=&-\left\langle TC_{\mathbf{k}\sigma }(\tau
)\overline{C}_{\mathbf{k} ^{\prime }\sigma ^{\prime }}(\tau
^{\prime })U(\beta )\right\rangle _{0}^{c},
\nonumber \\
&& \\
g_{\sigma \sigma ^{\prime }}(\tau |\tau ^{\prime })
&=&-\left\langle Tf_{\sigma }(\tau )\overline{f}_{\sigma ^{\prime
}}(\tau ^{\prime })U(\beta )\right\rangle _{0}^{c},  \nonumber
\label{1}
%
\end{eqnarray}
%
where index "$c$" means connected.

Fourier representation is denoted as $G_{\sigma \sigma ^{\prime
}}(\mathbf{k}, \mathbf{k}^{\prime }|i\omega )$ and $g_{\sigma
\sigma ^{\prime }}(i\omega )$ correspondingly. For these two
functions we have obtained the results$^{[1]}$ :
%
\begin{widetext}
%
\begin{eqnarray}
%
G_{\sigma }(\mathbf{k},\mathbf{k}^{\prime }|i\omega ) &=&\delta
_{\mathbf{kk} ^{\prime }}G_{\sigma }^{0}(\mathbf{k}|i\omega
)+\frac{V_{\mathbf{k}}V_{ \mathbf{k}^{\prime }}^{\ast
}}{N}G_{\sigma }^{0}(\mathbf{k}|i\omega )g_{\sigma }(i\omega
)G_{\sigma }^{0}(\mathbf{k}^{\prime }|i\omega ),
\nonumber \\
g_{\sigma }(i\omega ) &=&\frac{\Lambda _{\sigma }(i\omega )}{1-\Lambda
_{\sigma }(i\omega )G_{\sigma }^{0}(i\omega )}, \\
\Lambda _{\sigma }(i\omega ) &=&g_{\sigma }^{0}(i\omega
)+Z_{\sigma }(i\omega ),  \nonumber \label{2}
%
\end{eqnarray}
%
\end{widetext}
%
where zero order propagators of the conduction and impurity electrons have
the form ($\overline{\sigma }=-\sigma $).
%
\begin{eqnarray}
%
G_{\sigma }^{0}(\mathbf{k}|i\omega ) &=&(i\omega -\epsilon
(\mathbf{k}
))^{-1},  \nonumber \\
g_{\sigma }^{0}(i\omega ) &=&\frac{1-n_{\overline{\sigma }}}{i\omega
-\epsilon _{f}}+\frac{n_{\overline{\sigma }}}{i\omega -\epsilon _{f}-U},
\nonumber \\
n_{\overline{\sigma }} &=&\frac{\exp (-\beta \epsilon _{f})+\exp \left[
-\beta (2\epsilon _{f}+U\right] }{Z_{0}}, \\
Z_{0} &=&1+2\exp (-\beta \epsilon _{f})+\exp \left[ -\beta
(2\epsilon _{f}+U
\right] ,  \nonumber \\
G_{\sigma }^{0}(i\omega )
&=&\frac{1}{N}\sum\limits_{\mathbf{k}}|V_{
\mathbf{k}}|^{2}G_{\sigma }^{0}(\mathbf{k}|i\omega )=\int \frac{
V^{2}(\epsilon )\rho _{0}(\epsilon )d\epsilon }{i\omega -\epsilon
}. \nonumber \label{3}
%
\end{eqnarray}
%
Here $\epsilon (\mathbf{k})$ is the energy of conduction band and
$\epsilon _{f}$ of local impurity electrons, $\rho _{0}(\epsilon
)$ is the density of states of the bare conduction band and matrix
element of hybridization $V_{ \mathbf{k}}$ is supposed dependent
of the energy. $U$ is Coulomb repulsion of the impurity electrons.
$\omega \equiv \omega _{n}=(2n+1)\pi /\beta $ is odd Matsubara
frequency. The equations (2) are exact, but for correlation
function $ Z_{\sigma }(i\omega )$ doesn't exist exact Dyson type
equation and only the approximate contribution can be available:
see $Fig$. 9 of paper$^{[1]}$. Our main approximation formulated
in paper$^{[1]}$ comes to the summation of the ladder diagrams
which will be enough to obtain the main contributions of the spin
and charge fluctuations. This approximation has used only the
simplest irreducible Green's function $g_{2}^{(0)ir}$ which is
iterated many times. It has the form:
%
\begin{widetext}
%
\begin{eqnarray}
%
Z_{\sigma \sigma ^{\prime }}(\tau -\tau ^{^{\prime }}) & = &
-\sum\limits_{\mathbf{ k}_{1}\mathbf{k}_{2}\sigma _{1}\sigma
_{2}}\int\limits_{0}^{\beta }d\tau _{1}\int\limits_{0}^{\beta
}d\tau _{2}g_{2}^{(0)ir}[\sigma ,\tau ;\sigma _{1},\tau
_{1}|\sigma _{2},\tau _{2};\sigma ^{\prime },\tau ^{\prime
}]\frac{ 1}{N}V_{\mathbf{k}_{1}}^{\ast
}V_{\mathbf{k}_{2}}G_{\sigma _{2}\sigma _{1}}( \mathbf{k}_{2},\tau
_{2}|\mathbf{k}_{1},\tau _{1}),\label{4}
%
\end{eqnarray}
%
or in Fourier representation
%
\begin{eqnarray}
%
Z_{\sigma \sigma ^{\prime }}(i\omega )& = & -\frac{1}{\beta
}\sum\limits_{\omega _{1}}\sum\limits_{\sigma _{1}\sigma
_{2}}\sum\limits_{\mathbf{k}_{1}\mathbf{k
}_{2}}\widetilde{g}_{2}^{(0)ir}[\sigma ,i\omega ;\sigma
_{1},i\omega _{1}|\sigma _{2},i\omega _{1};\sigma ^{\prime
},i\omega ]\frac{1}{N}V_{ \mathbf{k}_{1}}^{\ast
}V_{\mathbf{k}_{2}}G_{\sigma _{2}\sigma _{1}}(\mathbf{k
}_{1},\mathbf{k}_{2}|i\omega _{1}).\label{5}
%
\end{eqnarray}
%
Here we take into account the conservation law of the frequencies:
%
\begin{eqnarray}
%
g_{2}^{(0)ir}[\sigma ,i\omega ;\sigma _{1},i\omega _{1}|\sigma
_{1},i\omega _{1};\sigma ^{\prime },i\omega ^{\prime }] & = &
\beta \delta (\omega -\omega ^{\prime
})\widetilde{g}_{2}^{(0)ir}[\sigma ,i\omega ;\sigma _{1},i\omega
_{1}|\sigma _{2},i\omega _{1};\sigma ^{\prime },i\omega
].\label{6}
%
\end{eqnarray}
%
In paramagnetic phase we have more simple equation ($\sigma
^{\prime }=\sigma $):
%
\begin{eqnarray}
%
Z_{\sigma }(i\omega ) & = & -\frac{1}{\beta }\sum\limits_{\omega
_{1}}\sum\limits_{\sigma _{1}}\widetilde{g}_{2}^{(0)ir}[\sigma
,i\omega ;\sigma _{1},i\omega _{1}|\sigma _{1},i\omega _{1};\sigma
,i\omega ]G_{\sigma _{1}}(i\omega _{1}),\label{7}
%
\end{eqnarray}
%
\end{widetext}
where
%
\begin{equation}
%
G_{\sigma }(i\omega
)=\frac{1}{N}\sum\limits_{\mathbf{k}_{1}\mathbf{k}
_{2}}V_{\mathbf{k}_{1}}^{\ast }V_{\mathbf{k}_{2}}G_{\sigma
}(\mathbf{k}_{1}, \mathbf{k}_{2}|i\omega ). \label{8}
%
\end{equation}
%
On the base of equations (2) and (3) the last function (8) can be
presented in the form:
%
\begin{equation}
%
G_{\sigma }(i\omega )=G_{\sigma }^{0}(i\omega )+[G_{\sigma
}^{0}(i\omega )]^{2}g_{\sigma }(i\omega )=\frac{G_{\sigma
}^{0}(i\omega )}{1-\Lambda _{\sigma }(i\omega )G_{\sigma
}^{0}(i\omega )}.\label{9}
%
\end{equation}
%
By using the definition (2) of correlation function of normal
state $\Lambda _{\sigma }(i\omega )$ and approximation (7) for
function $Z_{\sigma }(i\omega )$, we obtain the final integral
equation for $\Lambda _{\sigma }$:
%
\begin{widetext}
%
\begin{eqnarray}
%
\Lambda _{\sigma }(i\omega )=g_{\sigma }^{(0)}(i\omega
)-\frac{1}{\beta } \sum\limits_{\omega _{1}}\sum\limits_{\sigma
_{1}}\widetilde{g} _{2}^{(0)ir}[\sigma ,i\omega ;\sigma
_{1},i\omega _{1}|\sigma _{1,}i\omega _{1};\sigma ,i\omega
]\frac{G_{\sigma _{1}}^{0}(i\omega _{1})}{1-\Lambda _{\sigma
_{1}}(i\omega _{1})G_{\sigma _{1}}^{0}(i\omega _{1})}.\label{10}
%
\end{eqnarray}
%
\end{widetext}
%

In the second Section of this paper we shall discuss the simplest case of
symmetric impurity Anderson model with the condition $2\epsilon _{f}+U=0$.
In Section III the spectral function of the impurity electrons is analyzed and
the last Section IV contains the conclusions.
%

\section{Symmetric model}

%
In symmetric case when $\epsilon _{f}=-U/2<0$ and $\epsilon
_{f}+U=U/2>0$ we have more simple equations:
%
\begin{eqnarray}
%
g_{\sigma }^{0}(i\omega ) &=&\frac{i\omega }{(i\omega
)^{2}-(U/2)^{2}},
\nonumber \\
Z_{0} &=&2(1+\exp (-\beta \epsilon _{f}))=2(1+\exp (\beta U/2)),  \nonumber
\\
n_{\sigma } &=&1/2 \label{11}
%
\end{eqnarray}
%
and the antisymmetry property of zero order impurity Green's
function $ g_{\sigma }^{0}(-i\omega )=-g_{\sigma }^{0}(i\omega )$
takes place.

Additionally we suppose also the evenness of the matrix element $V(\epsilon
) $ and of the bare density of state $\rho _{0}(\epsilon )$. In this case
the function $G_{\sigma }^{0}(i\omega )$ is also antisymmetric $G_{\sigma
}^{0}(-i\omega )=-G_{\sigma }^{0}(i\omega )$. \newline
Thanks these antisymmetry properties we shall look for the antisymmetric
solution
%
\begin{equation}
%
\Lambda _{\sigma }(-i\omega )=-\Lambda _{\sigma }(i\omega
)\label{12}
%
\end{equation}
%
of the equation (10). Analytical continuation of these functions
have the property of oddness of their real parts and evenness of
imaginary parts in conformity with equations
%
\begin{eqnarray}
%
g_{\sigma }^{0}(E+i\delta ) &=&-g_{\sigma }^{0}(-E-i\delta ),
\nonumber
\\
G_{\sigma }^{0}(E+i\delta ) &=&-G_{\sigma }^{0}(-E-i\delta ), \\
\Lambda _{\sigma }^{0}(E+i\delta ) &=&-\Lambda _{\sigma
}^{0}(-E-i\delta ), \nonumber \label{13}
%
\end{eqnarray}
%
where
%
\begin{eqnarray}
%
G_{\sigma }^{0}(E+i\delta ) &=&I(E)-i\Gamma (E),  \nonumber \\
I(E) &=& P\int\frac{|V(\epsilon )|^{2}\rho _{0}(\epsilon
)d\epsilon }{E-\epsilon }
, \\
\Gamma (E) &=&\pi |V(E)|^{2}\rho _{0}(E).  \nonumber \label{14}
%
\end{eqnarray}
%
$I(E)$ is the principal part of the integral. This function is
antisymmetric. $\Gamma (E)$ is the band width of the virtual level
and an even function of energy. The symmetric impurity Anderson
model has the advantage to be of a simple form for the irreducible
two particles Green's functions in different spin and frequency
channels. In this special case we have$^{[2,3]}$
%
\begin{widetext}
%
\begin{eqnarray}
%
\widetilde{g}_{2}^{(0)ir}[\sigma ,i\omega ;\sigma ,i\omega
_{1}|\sigma ,i\omega _{1};\sigma ,i\omega ] & = & \frac{\beta
(U/2)^{2}(1-\delta \omega \omega _{1})}{[(i\omega
)^{2}-(U/2)^{2}][(i\omega _{1})^{2}-(U/2)^{2}]},\label{15}
%
\end{eqnarray}
%
\begin{eqnarray}
%
&&\widetilde{g}_{2}^{(0)ir}[\sigma ,i\omega ;\overline{\sigma
},i\omega _{1}| \overline{\sigma },i\omega _{1};\sigma ,i\omega
]=\frac{U}{2}\left\{ \frac{ \beta U}{2}\frac{(1-\exp (\beta
U/2))}{(1+\exp (\beta U/2))[\omega
^{2}+(U/2)^{2}][\omega _{1}^{2}+(U/2)^{2}]}-\right.  \nonumber \\
&&-\frac{\beta U\delta (\omega -\omega _{1})\exp (\beta
U/2)}{(1+\exp (\beta U/2))[\omega ^{2}+(U/2)^{2}][\omega
_{1}^{2}+(U/2)^{2}]}+\frac{\beta U\delta (\omega +\omega
_{1})}{(1+\exp (\beta U/2))[\omega ^{2}+(U/2)^{2}][\omega
_{1}^{2}+(U/2)^{2}]}- \nonumber \\
&&-\frac{2}{[\omega ^{2}+(U/2)^{2}][\omega
_{1}^{2}+(U/2)^{2}]}+4(U/2)^{2} \left[ \frac{1}{[\omega
^{2}+(U/2)^{2}][\omega _{1}^{2}+(U/2)^{2}]^{2}}+\right.
 \\
&&\left. \left. +\frac{1}{[\omega ^{2}+(U/2)^{2}]^{2}[\omega
_{1}^{2}+(U/2)^{2}]} \right] \right\} .  \nonumber \label{16}
%
\end{eqnarray}
%
Now we come back to equation (10) and note that thanks the
antisymmetry property (12) of $\Lambda _{\sigma }(i\omega )$
function only those terms of equations (15) and (16) which contain
Kronecker $\delta $ - symbols give the non zero contribution in
the right hand part of it. The result of summation has the form:
%
\begin{equation}
%
\Lambda _{\sigma }(i\omega )=\frac{i\omega }{(i\omega
)^{2}-(U/2)^{2}}+\frac{ 3(U/2)^{2}}{[(i\omega
)^{2}-(U/2)^{2}]^2}\frac{G_{\sigma }^{0}(i\omega )}{ [1-\Lambda
_{\sigma }(i\omega )G_{\sigma }^{0}(i\omega )]}. \label{17}
%
\end{equation}
%

We notice that the scattering channel with opposite spins (16)
gives in equation (17) the twice contribution in comparison with
parallel spin channel (15) and both of them are added together
giving the factor 3 in the right - hand part of equation (17).
There are two solutions of equation (17) and we take that of them
which has the correct asymptotic behavior $\lambda _{\sigma
}(i\omega )\rightarrow \frac{1}{i\omega }$ when $ |\omega |$ tends
to infinity. This solution has the form:
%
\begin{eqnarray}
%
\Lambda _{\sigma }(i\omega ) &=&\frac{1}{2G_{\sigma }^{0}(i\omega
)[(i\omega )^{2}-(U/2)^{2}]}\left\{ [(i\omega
)^{2}-(U/2)^{2}+i\omega G_{\sigma
}^{0}(i\omega )]-[(i\omega )^{2}-(U/2)^{2}-\right.   \nonumber \\
&-&\left. i\omega G_{\sigma }^{0}(i\omega )]\sqrt{1-12Q(i\omega
)}\right\} . \label{18}
%
\end{eqnarray}
%
\end{widetext}
where 
%
\begin{equation}
%
Q(i\omega )=\left( \frac{(U/2)G_{\sigma }^{0}(i\omega )}{(i\omega
)^{2}-(U/2)^{2}-i\omega G_{\sigma }^{0}(i\omega )}\right) ^{2}.
\label{19}
%
\end{equation}
%
We have used that branch of square root which gives one when
$Q(i\omega )$ tends to zero. On the base of equations (2) and (8)
we obtain the renormalized impurity electron propagator
\begin{widetext}
%
\begin{eqnarray}
%
g(i\omega )& = & \frac{2[g^{0}(i\omega )[G^{0}(i\omega
)]^{-1}+3Q(i\omega )(g^{0}(i\omega )-[G^{0}(i\omega
)]^{-1})^{2}]}{[[G^{0}(i\omega )]^{-1}-g^{0}(i\omega
)][1+\sqrt{1-12Q(i\omega )}+6Q(i\omega )(g^{0}(i\omega
)G^{0}(i\omega )-1)]}. \label{20}
%
\end{eqnarray}
%
\end{widetext}
%
In the last equations the spin index $\sigma $ is omitted because it is not
significant. Equation (20) has been obtained by taking into account the spin
and charge fluctuations contained in the correlation function $Z_{\sigma
}(i\omega )$.

The spectral function of the impurity electrons is equal to
%
\begin{equation}
%
A_{f}\left( E\right) =-2Img\left( {E+i\delta }\right) ,\label{21}
%
\end{equation}
%
where $g(E+i\delta )$ with $\delta =+0$ is the analytical
continuation of the Matsubara to retarded Green's function. In
absence of the correlation function $Z_{\sigma }(i\omega )$
instead of equation (20) a more simple form appears
%
\begin{equation}
%
g_{\sigma }^{I}(i\omega )=\frac{g_{\sigma }^{0}(i\omega
)}{1-g_{\sigma }^{0}(i\omega )G_{\sigma }^{0}(i\omega
)}=\frac{1}{\left[ g_{\sigma }^{0}(i\omega )\right]
^{-1}-G_{\sigma }^{0}(i\omega )},\label{22}
%
\end{equation}
%
which can be named as Hubbard I approximation. This equation
contains the zero order Green's function $g^{0}(i\omega )$
determined by equation (11) and averaged by hybridization
conduction electron function $G_{\sigma }^{0}(i\omega )$ of
equation (14).

Analytical continuation of equation (22) gives
%
\begin{equation}
%
g_{\sigma }^{I}(E+i\delta
)=\frac{1}{E-(U/2)^{2}E^{-1}-I(E)+i\Gamma (E)}.\label{23}
%
\end{equation}
%
We shall compare our renormalized spectral function (21) with more
simple
bare quantity $A_{f}^{0}(E)$: %
\begin{widetext}
%
\begin{eqnarray}
%
A_{f}^{0}(E)=-2Img^{0}(E+i\delta )& = & 2\pi \delta
(E-(U/2)^{2}E^{-1})=\pi \left( \delta (E-\frac{U}{2})+\delta
(E+\frac{U}{2})\right)\label{24}
%
\end{eqnarray}
%
and the value $A_{f}^{I}(E)$ obtained in Hubbard I approximation:
%
\begin{eqnarray}
%
A_{f}^{I}(E) & = & -2Img^{I}(E+i\delta )=\frac{2\Gamma (E)}{\left(
E-(U/2)^{2}E^{-1}-I(E)\right) ^{2}+\Gamma ^{2}(E)}.\label{25}
%
\end{eqnarray}
%
\end{widetext}
%
Here the odd $I(E)$ and even $\Gamma (E)$ functions are determined by
equation (14). Two resonances of equation (24) situated at energies $E=\pm
\frac{U}{2}$ have not the width. After some interactions taken into account
by Hubbard I approximation a new spectral function (25) appears. It has two
resonances with shifted values of energies $E=\pm E_{0}$ determined by the
presence of function $I(E)$:
%
\begin{equation}
%
E_{0}-\frac{(U/2)^{2}}{E_{0}}-I(E_{0})=0.\label{26}
%
\end{equation}
%
These resonances are broadened by the presence of the function
$\Gamma (E)$, which is the width of the virtual level. This
function $\Gamma (E)$ determines the height and width of the both
resonances. Near the new values of resonance energies $\pm E_{0}$
we can approximate (25) with more simple Lorentzian forms:
%
\begin{equation}
%
\frac{2\Gamma }{\psi ^{2}(E\mp E_{0})^{2}+\Gamma ^{2}},\label{27}
%
\end{equation}
%
where
%
\begin{equation}
%
\psi =1+(\frac{U}{2E_{0}})^{2}-I^{\prime }(E_{0})\label{28}
%
\end{equation}
%
and $E_{0}$ is determined by equation (26).

The both approximations (24) and (25) give the zero value of
spectral functions on the Fermi surface, where $E=0$. The
correctness of the result has to be verified by the sum rule
%
\begin{equation}
%
\int\limits_{-\infty }^{\infty }A_{f}(E)dE=2\pi .\label{29}
%
\end{equation}
%

Equation (24) fulfils this condition. If we omit for the simplicity function
$I(E)$ in equation (25) we can verify also the fulfillment of this condition
for equation (25), but not for its approximations (27), because the
parameter $\psi $ is not equal to two. Some details connected with the
choice of the density states can be found in Appendix.

In the next Section we use the equations (19) and (20) to obtain more
complete spectral function of impurity electrons and to verify the existence
of the resonance at zero energy.

\section{Renormalized spectral function}

Analytical continuation of equations (19) and (20) which are
necessary to calculate the spectral function have the form
\begin{widetext}
%
\begin{eqnarray}
%
g(E+i\delta ) &=&[[G^{0}(E+i\delta )]^{-1}-g^{0}(E+i\delta
)]^{-1}\times
\nonumber \\
&\times &\frac{2\left[ g^{0}(E+i\delta )[G^{0}(E+i\delta
)]^{-1}+3Q(E+i\delta )(g^{0}(E+i\delta )-[G^{0}(E+i\delta
)]^{-1})^{2}\right] }{\left[ 1+\sqrt{ 1-12Q(E+i\delta )}+6Q(E+i\delta
)(g^{0}(E+i\delta )G^{0}(E+i\delta )-1) \right] },\label{30}
%
\end{eqnarray}
%
\end{widetext}
%
\begin{equation}
%
Q(E+i\delta )=\left(\frac{UG^{0}(E+i\delta
)/2}{E^{2}-(U/2)^{2}-EG^{0}(E+i\delta )}\right)
^{2}.\label{31}
%
\end{equation}
%
First of all we shall analyze the behavior of the renormalized Green's
function on the Fermi surface for $E=0$. Near the value of the energy $E=0$
we can approximate the quantity $Q(E+i\delta )$ by the expression
%
\begin{equation}
%
Q(E+i\delta )\mid _{E=0}=-\left( \frac{2\Gamma (0)}{U}\right)
^{2}\label{32}
%
\end{equation}
%
and supposing the smallness of the parameter $2\Gamma /U$\ \ we
can approximate equation (30) by more simple one:
%
\begin{widetext}
%
\begin{eqnarray}
%
g(E+i\delta )& \simeq & \frac{6Q(E+i\delta )}{G^{0}(E+i\delta
)[1+\sqrt{ 1-12Q(E+i\delta )}-6Q(E+i\delta )]}.\label{33}
%
\end{eqnarray}
%
For little values of energy we have
%
\begin{eqnarray}
%
g(E+i\delta )& \simeq & \frac{6\left( 2\Gamma /U\right)
^{2}}{\Gamma ^{2}(E)+E^{2}I^{\prime }(0)^{2}}\frac{1}{\left[
1+\sqrt{1+12\left( 2\Gamma /U\right) ^{2}}+6\left( 2\Gamma
/U\right) ^{2}\right] }\left[ -EI^{\prime }(0)+i(\Gamma
(E)+\frac{I^{\prime }(0)^{2}E^{2}}{\Gamma (0)})\right] .\label{34}
%
\end{eqnarray}
%

The spectral function of impurity electrons in the region of
little values of energy $E$ is different from zero and has the
form
%
\begin{eqnarray}
%
A_{f}(E)\ & \simeq & \frac{12\left( 2\Gamma /U\right) ^{2}/\Gamma
(0)}{\left[ 1+6\left( 2\Gamma /U\right) ^{2}+\sqrt{ 1+12\left(
2\Gamma /U\right) ^{2}}\right] }.\label{35}
%
\end{eqnarray}
%
\end{widetext}
%
In Appendix are cited the values of the quantity $I^{\prime }(0)$
equal to $4\rho (0)V^{2}(0) $/$W$ or $\pi V^{2}(0)\rho (0)/D$ for
two different model of density of states. The result (35) differs
essentially from the zero value obtained in such more simple
approximations for $A_{f}(E)$ as $A_{f}^{0}(E)$ and
$A_{f}^{I}(E)$. Thus we have established the existence of two
resonances at energies $E=\pm E_{0}$ determined by (26) and the
peculiarity at $E=0$. We can find the corrections for spectral
function $A_{f}(E)$ for two
resonances $E=\pm E_{0}. $ With this end in view we determine the quantity $%
Q $ for $E=E_{0}$
%
\begin{equation}
%
Q(E+i\delta )=-\left( I(E_{0})-i\Gamma (E_{0})\right) ^{2}\left(
\frac{U}{ 2E_{0}\Gamma (E_{0})}\right) ^{2},  \nonumber \label{36}
%
\end{equation}
%
where we suppose that the value $E_{0}$ is inside the edges of the
band
width. After some calculations made with the supposition that quantities $%
I(E_{0})$/$\Gamma (E_{0})$ and $UI(E_{0})/2E_{0}\Gamma (E_{0})$
are large and are superior in number to one we obtain
%
\begin{equation}
%
A_{f}(E_{0})=\frac{1}{\sqrt{3}\left( 1+3(U/2E_{0})^{2}\right)
}\cdot \frac{ 4E_{0}}{UI(E_{0})}.\label{37}
%
\end{equation}
%
This quantity essentially differs from the value
$A_{f}^{I}(E_{0})=2/\Gamma (E_{0})$ being considerable less than
it. Thus the process of renormalization of the impurity electron
propagator results in appearance of the peculiarity at $E=0$ and
to diminution of two resonances situated at $E=\pm E_{0}$.

\begin{figure}[!h]
%
\centering
\includegraphics[width=0.45\textwidth,clip]{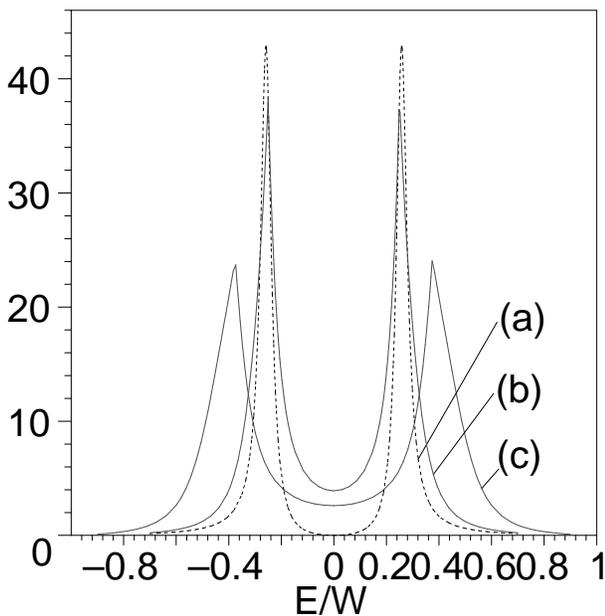}
\caption{Spectral function $A_{f}(E)\times W$ for different values
of the theory parameters as function of energy $E/W$ in the
Hubbard I approximation (case (a)) and in the our ladder
approximation (cases (b) and (c)). In the cases (a) and (b)
$U=1\,eV,\quad W=2\,eV,\quad \Gamma =0.1\,eV $ and in case case
(c) $U=1.5\,eV,\quad W=2\,eV,\quad \Gamma =0.15\,eV $. }
\label{fig-1}
%
\end{figure}

The full spectral function can can be calculated on the base of
the equation (30) and (31). On the Figure \ref{fig-1} the results
of numerical investigation of the full spectral function are
presented for different values of theory parameters. For
comparison on this figure also the result obtained in Hubbard I
approximation is presented.
%
%
As can be seen from Figure \ref{fig-1} there are two sharp
resonance peaks near the energies $E=\pm E_0$ and smooth behaviour
near E=0.
 The distance between peaks is determined by parameter $U$ and the
 height and width of peaks are determine by parameter $\Gamma$.

\section{Conclusions}

We discuss the symmetric Anderson impurity model and take into account the
strong electronic correlations of the impurity electrons by elaborating the
suitable diagram technique. This paper is based on the previous our one$%
^{[1]}$ founded on the diagrammatical investigation and analysis
of the properties of single-site Anderson model. First of all we
establish the antisymmetry property of the functions
$g^{0}(i\omega )$, $\Lambda _{\sigma }(i\omega )$ and $g_{\sigma
}(i\omega )$, and use the exact Dyson type equations for our
functions. The special approximation for the correlation function
$Z_{\sigma }(i\omega )$ has been obtained which gives the
possibility to close the system of equations and to find the
solution for renormalized function $g_{\sigma }(i\omega )$. This
Matsubara Green's function is continued analytically to obtain the
retarded one.

Spectral function of impurity electrons is obtained
and the structure of resonances
and their properties are analyzed. Two of resonances of this function at $%
E=\pm E_{0}$ (Figure \ref{fig-1}) correspond to the energies of
quantum transitions of single --
site impurity and the smooth behaviour was found at the energy $%
E=0$. The details of the spectral function renormalization are based on the
properties of real $I(E)$ and imaginary $\Gamma (E)$ parts of function $%
G_{\sigma }^{0}(i\omega )$, which is conduction band electron function
averaged by the hybridization interaction. The values of these functions and
of the values of energies $\pm E_{0}$ are discussed in a special Appendix.


\begin{acknowledgments}

This work was supported by the Heisenberg -- Landau Program. It is
a pleasure to acknowledge the discussions with Professor N.M.
Plakida. V.A. M. would like to thank the Duisburg - Essen
Univeristy for financial support
and hospitality. %

\end{acknowledgments}

%

%
\appendix
%

\section{Simple examples of density of states}

%
We can demonstrate some simple examples of the choice of the
density of states and of the corresponding functions $I(E)$ and
$\Gamma (E)$. For simplicity the energy dependence of the matrix
element of hybridization $ V(\epsilon )$ is supposed smooth and is
neglected. One example of density of states has been proposed in
the paper$^{[4]}$. In this paper the following equations are used
%
\begin{eqnarray}
%
\rho _{0}(\epsilon ) &=&\rho _{0}(0)\left( 1-\left( \epsilon
/U\right)
^{2}\right) ,\qquad \left\vert \epsilon \right\vert <W,  \nonumber \\
I(\epsilon ) &=&\rho _{0}(0)V^{2}(0)[2\epsilon /W+\left( \epsilon
^{2}/W^{2}-1\right) \ln \left\vert \frac{\epsilon -W}{\epsilon +W}
\right\vert] ,  \nonumber \\
\Gamma (\epsilon ) &=&\pi V^{2}(\epsilon )\rho _{0}(\epsilon
).\label{A1}
%
\end{eqnarray}
%
where $2W$ is the conduction band width. For little value of
energy we have $ I(\epsilon )=I^{\prime }(0)\epsilon $ \ with
%
\begin{equation}
%
I^{\prime }(0)=4\rho _{0}(0)V^{2}(0)/W \label{A2}
%
\end{equation}
%
and for $E\rightarrow \pm \infty $ function $I(E)$ tends to zero
as $1/E$. In the case $(A1)$ the equation (26) takes the form
$\left( x_{0}=E_{0}/W\right) $:
%
\begin{equation}
%
x_{0}-a^{2}/x_{0}-b\varphi (x_{0})=0,\label{A3}
%
\end{equation}
%
where
%
\begin{eqnarray}
%
\varphi (x) &=&2x+(x^{2}-1)\ln \left\vert
\frac{x-1}{x+1}\right\vert
;\,\,a=U/2W;\,\,  \nonumber \\
b &=&\rho (0)V^{2}/W.  \nonumber
%
\end{eqnarray}
%
We note that the functions $\rho _{0}(\epsilon )$ and $\Gamma
(\epsilon )$ exist only inside the edges of the conduction
electron band $\left\vert E\right\vert <W$ whereas the function
$I(\epsilon )$ and the solution of equation $(A2)$ can exist also
for $\left\vert E\right\vert >W$. Therefore we have to find the
solution of $(A2)$ with $x<1$ and consider the conditions for the
values of parameters $a$ and $b$ compatible with this requirement.

Another simple example of the density of state is one with
Lorentzian shape$ ^{[5]}$
%
\begin{equation}
%
\rho _{0}(\epsilon )=2D/(E^{2}+D^{2}),\label{A4}
%
\end{equation}
%
with chemical potential placed at the $\epsilon =0$. This choice
has the advantage of not introducing band edges. It has the
parameter $D$ as an effective band width:
%
\begin{eqnarray}
%
I(E) &=&2\pi V^{2}(0)E/(E^{2}+D^{2})  \nonumber \\
&& \\
\Gamma (E) &=&\pi \rho _{0}(0)V^{2}(0)=2\pi V^{2}(0)/D.  \nonumber
\label{A5}
%
\end{eqnarray}
%
In this case instead of equation $(A3)$ we have other values of
parameters $ a$ and $b$ and other form of function $\varphi (x)$:
%
\begin{eqnarray}
%
x-a^{2}/x-bx/(1+x^{2}) &=&0,  \nonumber \\
&& \\
a=U/2D;\,\,\,b &=&2\pi V^{2}(0)/D^{2}.  \nonumber \label{A6}
%
\end{eqnarray}
%
Equation ($A6)$ has two solutions $\pm x_{0}$ with
%
\begin{equation}
%
x_{0}=\left[ \sqrt{(1-a^{2}+b)^{2}+4a^{2}}+a^{2}+b-1\right]
^{1/2}/\sqrt{2}.\label{A7}
%
\end{equation}
%
The value of parameter $I^{\prime }(0)$ is equal to $2\pi
V^{2}(0)/D^{2}=\pi V^{2}(0)\rho _{0}(0)/D.$
%

\end{document}